\documentclass[runningheads]{llncs}
\usepackage[T1]{fontenc}
\usepackage{graphicx}
\usepackage{algorithm}
\usepackage{algorithmic}
\usepackage{listings}
\usepackage{xcolor}
\usepackage{url}
\usepackage{float}
\usepackage{array} 
\usepackage{amsmath}
\usepackage{amsfonts}
\usepackage{longtable}
\usepackage{colortbl}
\usepackage{booktabs}
\usepackage{url}

\usepackage{subcaption}  

\usepackage{tikz}
\usetikzlibrary{shapes.geometric, arrows.meta, positioning, calc}

\usepackage{multirow}

\newcommand{\PHALYNX}{{\tt Spinel}}
\newcommand{\spinel}{\PHALYNX}
\newcommand{\SPHINCSPLUS}{{\tt SPHINCS+}}
\newcommand{\WOTSPLUS}{{\tt WOTS+}}

\newcommand{\SPHINCS}{{\tt SPHINCS}}
\newcommand{\MSS}{{\tt MSS}}

\newcommand{\HORS}{{\tt HORS}}
\newcommand{\HORST}{{\tt HORST}}
\newcommand{\FORS}{{\tt FORS}}

\newcommand{\SHA}{{\tt SHA-256}}
\newcommand{\SHAKE}{{\tt SHAKE-256}}
\newcommand{\Haraka}{{\tt Haraka}}

\begin{document}

\title{\PHALYNX: A Post-Quantum Signature Scheme Based on $\operatorname{SL}_n(\mathbb{F}_p)$ Hashing}

\author{Asmaa~Cherkaoui\inst{1} \and
Faraz~Heravi\inst{2} \and 
Delaram~Kahrobaei\inst{3,4,5,6} \and
Siamak~F.~Shahandashti\inst{6} 
}
\authorrunning{Cherkaoui et al.}



\institute{
Laboratory of Mathematical Analysis, Algebra and Applications (LAM2A), \\ Faculty of Sciences Ain Chock (FSAC), Hassan II University, Casablanca, Morocco \\
\email{esma1maysan@gmail.com} 
\and 
Department of Computer Science, University of Texas, Austin, USA\\
\email{farazh@utexas.edu}
\and
Departments of Computer Science and Mathematics, Queens College, \\ 
City University of New York, USA\\
\email{delaram.kahrobaei@qc.cuny.edu} 
\and
PhD Program in Mathematics, and Initiative for the Theoretical Sciences, Graduate Center, City University of New York, USA\\
\and
Department of Computer Science and Engineering, Tandon School of Engineering, New York 
University, USA\\
\and
Department of Computer Science, University of York, United Kingdom \\
\email{siamak.shahandashti@york.ac.uk} 
}

\maketitle

\begin{abstract}
The advent of quantum computation compels the cryptographic community to design digital signature schemes whose security extends beyond the classical hardness assumptions. 
In this work, we introduce \PHALYNX{}, a post-quantum digital signature scheme that combines the proven security of \SPHINCSPLUS{} (CCS~2019) with a new family of algebraic hash functions (Adv.\ Math.\ Commun.~2025) derived from the Tillich–Zémor paradigm (Eurocrypt~2008) with security rooted in the hardness of navigating expander graphs over $\operatorname{SL}_n(\mathbb{F}_p)$, a problem believed to be hard even for quantum adversaries. 
We first provide empirical evidence of the security of this hash function, complementing the original theoretical analysis. 
We then show how the hash function can be integrated within the \SPHINCSPLUS{} framework to give a secure signature scheme. We then model and analyze the security degradation of the proposed scheme, which informs the parameter selection we discuss next. Finally, we provide an implementation of the hash function and the proposed signature scheme \PHALYNX{} as well as detailed empirical results for the performance of \PHALYNX{} showing its feasibility in practice. 
Our approach lays the foundations for the design of algebraic hash-based signature schemes, expanding the toolkit of post-quantum cryptography.


\end{abstract}
\section{Introduction}

Digital signatures remain a cornerstone of modern cryptography, ensuring authenticity, integrity, and non-repudiation in digital communications. The touted imminent arrival of large-scale quantum computers threatens the security of widely-deployed signature schemes with security based on factoring and discrete logarithm, driving the urgent development of quantum-resistant alternatives.

Among the most prominent candidates are hash-based signature schemes, whose security reduces to the relatively well-understood properties of cryptographic hash functions. \SPHINCSPLUS{}~\cite{BernsteinHKNRS19sphincs+} is the leading stateless hash-based signature scheme, standardized through the NIST Post-Quantum Cryptography Standardization process~\cite{NISTPQC} as SLH-DSA in FIPS~205~\cite{slh-dsa-2024}. Its modular design, which integrates Winternitz one-time signatures (\WOTSPLUS{})~\cite{Hulsing13wots+}, Forest of Random Subset (\FORS{}), and hypertrees, enables the provision of strong security guarantees. Being a hash-based signature, \SPHINCSPLUS{}'s security also crucially relies on that of the underlying hash functions such as \SHA{} and \SHAKE{} whose security is primarily established heuristically. The security of algebraically structured hash functions, on the other hand, is based on well-understood and extensively studied mathematical problems. Hence, a natural question arises: can algebraic hash functions strengthen the theoretical underpinnings of hash-based signature schemes and diversify the design landscape in terms of the relied-upon security assumptions? 

Recent research has significantly advanced the practicality of \SPHINCSPLUS{}. Formal verification has strengthened confidence in its security proof~\cite{TightSec}, while engineering optimizations have improved performance in real-world settings such as FIDO authentication on embedded devices~\cite{Fido}. Other works focus on reducing signature size through careful message preimage selection~\cite{SphincsPlusCompressed} or by refining parameter selection to balance bandwidth, speed, and overuse resilience~\cite{KreyviumSphincsSmaller}. Structural generalizations of the hypertree and one-time signature layers have further expanded the design space, enabling more favorable trade-offs between efficiency and compactness~\cite{ExtendingSphincsPlus}. However, all these approaches retain the original hash primitives (\SHA{}, \SHAKE{}, or \Haraka{}), whose security remains heuristic. None explore the replacement of the underlying hash function with a post-quantum primitive based on hard mathematical problems, an opportunity to strengthen the theoretical foundation of the security of \SPHINCSPLUS{}. 
In this work, we take this direction. Our starting point is recent work by Le~Coz et al.\ that introduced a family of post-quantum hash functions based on special linear groups $\operatorname{SL}_n(\mathbb{F}_p)$ \cite{LeCoz25pqhash}. This construction, rooted in the Tillich–Zémor paradigm \cite{TillichZemor94,TillichZemor08}, relies on random walks on Cayley graphs of matrix groups. The mathematical analysis in \cite{LeCoz25pqhash} demonstrates that these graphs exhibit rapid mixing, high girth, and spectral expansion, properties that imply strong resistance against collision and preimage attacks and excellent output randomness properties. 

We build on these results and introduce \PHALYNX{} (\emph{\textbf{S}$\mathrm{L}_4$-Based \textbf{P}ost-Quantum \textbf{I}ntegrity from \textbf{N}on-Backtracking \textbf{E}xpanders over \textbf{L}inear Groups}), a signature scheme that integrates the $\operatorname{SL}_n(\mathbb{F}_p)$-based hash of Le~Coz et al.\ into the \SPHINCSPLUS{} framework. 
To this end, we provide the following contributions: 
\begin{itemize}
    \item We implement the Le~Coz et al.\ hash function and demonstrate that it passes all the 15 categories of the  standard NIST Statistical Test Suite, thereby providing the first empirical evidence of its security, specially in terms of output randomness, complementing the previously established theoretical analyses. 
    \item We show how the \SPHINCSPLUS{} framework can be instantiated with an adapted version of the Le~Coz et al.\ hash function as its underlying primitive, and provide a detailed analysis of its security degradation informing the parameter selection for the designed signature to achieve practical security and efficiency. 
    \item We implement the proposed signature scheme and provide detailed empirical results, benchmarking its computational efficiency and signature and key lengths for a range of parameters from the parameter space. 
\end{itemize}


The remainder of the paper is organized as follows. 
Section~\ref{sec:sphincs-overview} revisits the \SPHINCSPLUS{} framework. 
Section~\ref{sec:hash} presents our adaptation of the Le~Coz et al.\ hash function to enable the design of \spinel. 
Section~\ref{sec:hash-evaluation} provides the results of our empirical evaluation of the hash function based on NIST statistical tests. 
Section~\ref{sec:phalynx-design} presents the \PHALYNX{} design and discusses parameter selection for the scheme. 
Section~\ref{sec:performance} provides the results of our empirical evaluation of \PHALYNX{}. 
Finally, we conclude with future directions for research. 

\section{Overview of \SPHINCSPLUS{} and Its Variants}
\label{sec:sphincs-overview}

\SPHINCSPLUS{}~\cite{BernsteinHKNRS19sphincs+} is the state-of-the-art stateless hash-based signature scheme, selected in the NIST Post-Quantum Cryptography standardization process. It inherits the overall design of its predecessor \SPHINCS{}~\cite{Bernstein15SPHINCS}, while introducing several modifications that improve both efficiency and security. Among these are the replacement of the underlying few-time signature \HORS{} with the more compact \FORS{}, a deterministic leaf-index selection based on the message and public seed, and the adoption of tweakable hash functions to strengthen resistance against multi-target attacks.

Importantly, \SPHINCSPLUS{} is not a single scheme, but a family of instantiations. Three hash primitives are supported: \SHA{}, \SHAKE{}, and \Haraka{}, each offering distinct trade-offs in terms of performance and implementation cost. Each primitive is paired with parameter sets targeting NIST security categories I, III, and V, and for every security level there exist both a \emph{fast} (f) and a \emph{small} (s) variant, balancing signature size against computational efficiency. For instance, the parameter set \SPHINCSPLUS{}-\SHAKE{}-128s achieves category I security with smaller signatures, whereas \SPHINCSPLUS{}-\SHAKE{}-128f prioritizes faster signing and verification.



To set the stage for \PHALYNX{}, which integrates a new group-theoretic hash into this framework, we next recall the main building blocks of \SPHINCSPLUS{}: the Winternitz one-time signature scheme (\WOTSPLUS{}), the few-time signature scheme \FORS{}, and the hypertree construction that enables stateless operation.

\subsection{\WOTSPLUS{}: Winternitz One Time Signature Scheme}

The Winternitz one-time signature scheme (\WOTSPLUS{}) \cite{Hulsing13wots+} is the foundational 
building block of \SPHINCSPLUS{}. It is designed for single-message signing, since reusing a private key 
across multiple messages compromises security. \WOTSPLUS{} balances efficiency and signature 
size through the choice of the \emph{Winternitz parameter} \(w > 1\).

The scheme depends on two main parameters: the security parameter \(n\), which determines the 
bit-length of keys, signatures, and hash outputs, and the Winternitz parameter \(w\). From these, 
two derived values are computed:
\[
\ell_1 = \left\lceil \frac{n}{\log w} \right\rceil, \qquad 
\ell_2 = \left\lfloor \frac{\log\!\left(\ell_1(w-1)\right)}{\log w} \right\rfloor + 1,
\]
and we set \(\ell = \ell_1 + \ell_2\). Thus, a \WOTSPLUS{} private key, public key, and signature 
each consist of \(\ell\) strings of length \(n\).

At the core of \WOTSPLUS{} lies a chaining function based on a hash \(H : \{0,1\}^n \to \{0,1\}^n\). 
Given an input \(x \in \{0,1\}^n\) and a sequence of bitmasks \(r=(r_1,\ldots,r_{w-1})\), the chain is 
defined recursively by
\[
c^0(x,r) = x, \qquad 
c^i(x,r) = H\big(c^{i-1}(x,r) \oplus r_i\big), \quad i = 1, ..., w-1.
\]
In practice (e.g., in \SPHINCSPLUS{}), the bitmasks are derived deterministically from a public seed and address using a pseudorandom function, ensuring stateless operation.
The private key consists of \(\ell\) random \(n\)-bit strings. The public key is obtained by applying the chaining function \(w-1\) times to each secret element. 
To sign a message \(M\), it is first represented in base \(w\), together with a checksum that prevents forgery through partial key exposure. The resulting vector \((b_1,\ldots,b_\ell)\) determines how many iterations of the chaining function are applied to each secret key element. 
The verifier applies the remaining $w-1-b_i$ iterations to each signature element so that every component is hashed exactly \(w-1\) times, and checks that the resulting vector matches the public key. 
%
The security of \WOTSPLUS{} reduces to the second-preimage resistance of the underlying hash function, assuming the bitmasks are unpredictable to the adversary. 


\subsection{\MSS{}: Merkle Signature Scheme}

While \WOTSPLUS{} provides strong security guarantees, it is limited to a single use of each key pair. To enable many signatures under one public key, one can organize a collection of one-time signature (OTS) keys into a binary Merkle tree~\cite{Merkle89}, resulting in a Merkle Signature Scheme (\MSS{}). A tree of height $h$ contains $2^h$ leaves, each corresponding to the hash of an OTS public key. The root of the tree serves as a compact master public key, thereby enabling scalable authentication of multiple signatures. 

The construction proceeds as follows. First, $2^h$ independent OTS key pairs are generated. 
Each OTS public key is hashed to form a leaf, and the internal nodes of the tree are computed 
recursively by hashing the concatenation of their children. 
This process culminates in the root node, which acts as the public key of the \MSS{}, while the 
private key consists of all underlying OTS secret keys.

To sign a message, the signer selects an unused OTS key pair and generates a \WOTSPLUS{} (or 
similar OTS) signature. The \MSS{} signature is then augmented with an \emph{authentication path}: 
a sequence of sibling nodes that allows the verifier to recompute the path from the chosen leaf 
to the root. Verification involves checking the validity of the OTS signature and confirming that 
the reconstructed root matches the known \MSS{} public key.

\MSS{} provides a generic method to extend one-time signatures into many-time signatures while retaining compact public keys. Its logarithmic authentication path ensures that verification remains efficient even for large numbers of leaves. In modern hash-based schemes like \SPHINCSPLUS{}, this idea is generalized into a hypertree structure composed of multiple stacked Merkle trees, enabling practical support for $2^{64}$ or more signatures. Moreover, \SPHINCSPLUS{} employs tweakable hash functions, where each hash call is domain-separated via a unique address, to prevent multi-target and cross-tree attacks. The security of \MSS{} (and its extensions) reduces to the collision resistance of the underlying hash function and the security of the OTS scheme, making the choice of hash primitive a critical design decision.

\subsection{\FORS{}: Forest of Random Subsets}

Early hash-based signature schemes sought to extend one-time signatures (OTS) to support multiple messages without resorting to a full Merkle tree per signature. The Hash to Obtain Random Subset (\HORS{}) scheme~\cite{ReyzinR02hors} introduced the few-time signature (FTS) paradigm, allowing a bounded number of signatures by revealing indexed secret values. However, \HORS{} suffers from progressive key exposure: each signature leaks secret components, weakening security with reuse.

The \SPHINCS{} design~\cite{Bernstein15SPHINCS} improved upon \HORS{} with \HORST{} (Hash to Obtain Random SubTrees), which organizes secret values into a single Merkle tree, reducing public key size and enabling authentication paths. Yet \HORST{} still exhibits significant leakage across signatures and lacks tight security bounds.

In \SPHINCSPLUS{}, both are superseded by \FORS{} (Forest of Random Subsets) \cite{BernsteinHKNRS19sphincs+}, which replaces the monolithic tree with a forest of independent trees, offering stronger security, better parameter flexibility, and tighter reductions to standard hash function security assumptions.



\FORS{} is parameterized by two integers: 
$k$, the number of trees, and 
$t = 2^a$, the number of leaves per tree (with $a$ being the tree height).
A message digest of $m = k a$ bits is parsed into $k$ indices $(M_1, \dots, M_k)$, where each $M_i \in \{0, \dots, t-1\}$ selects a leaf in the $i$-th tree.

The secret key consists of $k t$ strings, each $n$-bit long, derived deterministically from a master secret seed using a pseudorandom function (PRF), ensuring stateless key generation. For each tree $i$, the $t$ secret values form the leaves of a binary Merkle tree of height $a$. The root of each tree is computed using a tweakable hash function $H$, with unique domain separators (e.g., address bits) per node to prevent collisions across trees and layers. The public key is then obtained by compressing the $k$ roots via another invocation of the tweakable hash: 
\(
\mathsf{pk}^{\mathrm{FORS}} = T_h\bigl(\mathsf{roots}^{(1)}, \dots, \mathsf{roots}^{(k)}\bigr)
\), 
where $T_h$ denotes a hash function instantiated with a distinct tweak to ensure domain separation.

\begin{figure}[tb]
    \centering
    \includegraphics[width=0.8\textwidth]{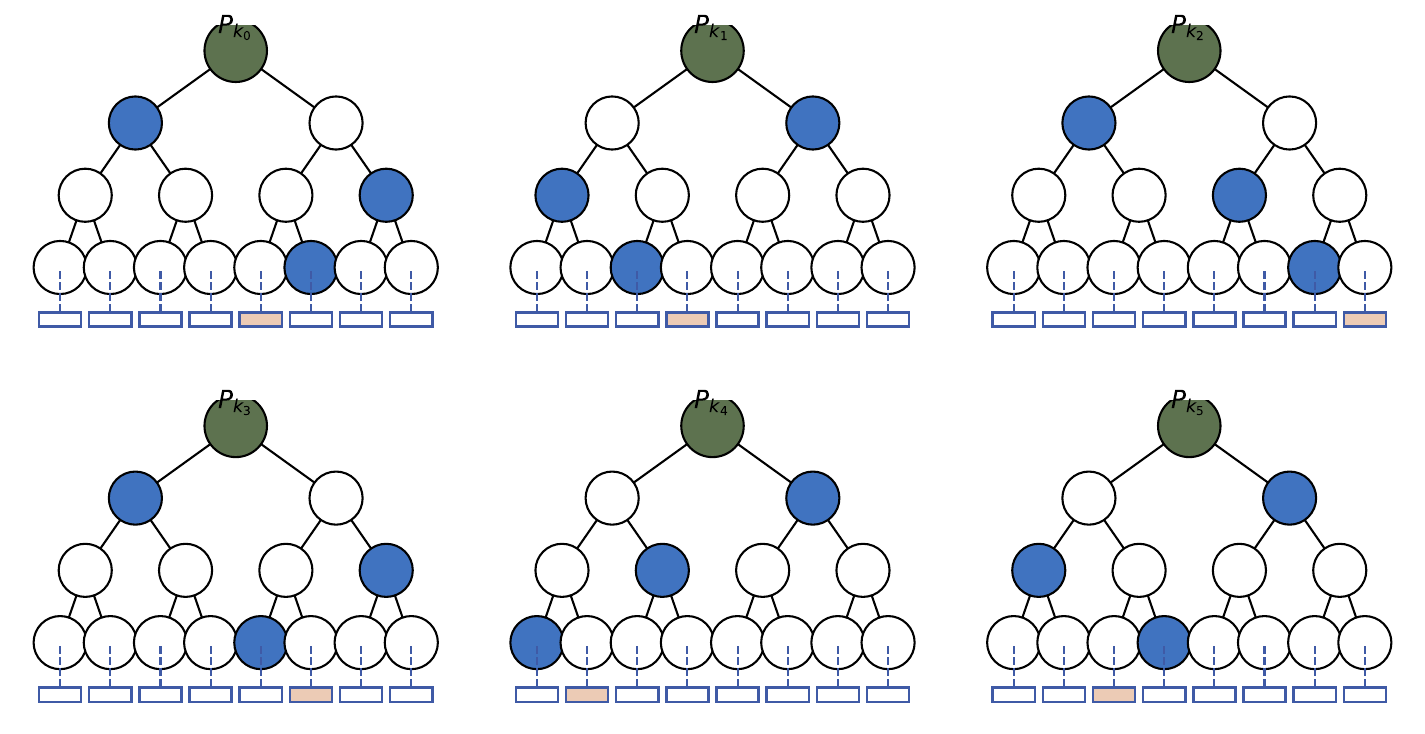}
    \caption{\FORS{} signing with $k=6$ trees, $t=2^a=8$ leaves per tree, and message digest parsed as $100\;011\;111\;101\;001\;010$. For each tree, one leaf (in red) is revealed along with its authentication path (in blue).}
    \label{fig:fors-signing}
\end{figure}

To sign a message, \FORS{} outputs 
the $k$ secret leaf values $\{s_{i,M_i}\}_{i=1}^k$, and 
the corresponding authentication paths (sibling nodes) for each leaf in its tree.
This is shown in Fig.~\ref{fig:fors-signing}. 
Verification recomputes each tree root from the revealed leaf and its authentication path, then applies $T_h$ to obtain the candidate public key, which is compared against the known value.


The forest structure ensures that even if two messages share identical indices in some positions, the secret values and authentication paths remain independent across trees. This isolation dramatically reduces the success probability of forgery under adaptive chosen-message attacks and enables tight security reductions to the collision resistance of the underlying hash function~\cite{SphincsPlus}. Consequently, \FORS{} strikes an optimal balance between signature size, signing cost, and provable security, making it a cornerstone of the \SPHINCSPLUS{} design.

\subsection{\SPHINCSPLUS{}}

A central challenge in hash-based signatures is scalability: one-time and few-time primitives such as \WOTSPLUS{} and \FORS{} can only sign a limited number of messages per key pair. To enable stateless, long-term use with an exponentially large signing capacity, \SPHINCSPLUS{} employs the \emph{hypertree}, a layered composition of Merkle trees that aggregates many \FORS{} instances under a single public key~\cite{BernsteinHKNRS19sphincs+}.

The hypertree is parameterized by its total height $h$ and the number of layers $d$, where $d$ divides $h$. Each layer $\ell \in \{0, \dots, d-1\}$ consists of a forest of Merkle trees of height $h/d$. The leaves of layer $\ell$ store the public keys of the signing instances at layer $\ell-1$ (with layer $0$ containing \FORS{} public keys). The root of the topmost tree (layer $d-1$) serves as the global public key of the scheme. This recursive structure yields a total capacity of $2^h$ distinct signatures, while keeping public keys compact and operations stateless.

During signing, a message is first signed using a fresh \FORS{} instance, yielding a \FORS{} signature and public key. The \FORS{} public key is then authenticated via a \WOTSPLUS{} signature under a leaf of the lowest Merkle tree (layer~0). The \WOTSPLUS{} public key is in turn used as a leaf in layer~1, and so on, up to the top layer. At each layer, an authentication path of length $h/d$ is revealed, linking the current node to its parent root. Crucially, all hash computations use \emph{tweakable hashing} with unique addresses to ensure domain separation across layers, trees, and nodes.

Verification reconstructs this chain bottom-up: it verifies the \FORS{} signature, recomputes the \WOTSPLUS{} public key, validates the Merkle authentication paths layer by layer, and confirms that the final root matches the global public key.




\SPHINCSPLUS{} avoids the prohibitive key material of na\"{i}ve Merkle trees by using a \emph{virtual} hypertree: only the signing path is instantiated on-demand. A message is first signed with a fresh \FORS{} instance (derived deterministically from a master seed), producing a \FORS{} signature and public key. This \FORS{} public key is then authenticated through a chain of \WOTSPLUS{} signatures, each embedded in a Merkle tree at successive layers of the hypertree. The root of the topmost layer serves as the global public key $\mathsf{PK}_{\text{root}}$.

Crucially, every hash invocation in \SPHINCSPLUS{}, whether in \WOTSPLUS{} chaining, \FORS{} leaf hashing, Merkle node computation, or digest derivation, is performed via a \emph{tweakable hash function}:
\[
T_h(P, \mathsf{ADRS}, M) = H(P \,\|\, \mathsf{ADRS} \,\|\, M),
\]
where $\mathsf{ADRS}$ is a structured 32-byte address encoding context (e.g., layer, tree index, leaf index, operation type). This ensures \emph{domain separation}: identical inputs in different contexts yield independent outputs, which is essential for security against multi-target attacks and for the tight reductions in the (quantum) random oracle model~\cite{BernsteinHKNRS19sphincs+}. The precise structure of the address ($\mathsf{ADRS}$) and its role in tweakable hashing are detailed in Appendix~\ref{sec:adrs}.

The private key consists of two $n$-bit seeds $(\mathsf{SK}_1, \mathsf{SK}_2)$ and a set of bitmasks $Q$ (or, in the \SHAKE{}/ \Haraka{} variants, a PRF-derived mask). The seed $\mathsf{SK}_1$ deterministically generates all signing keys (\FORS{} and \WOTSPLUS{}), while $\mathsf{SK}_2$ provides message-dependent randomness for index selection and digest computation. The public key is $(\mathsf{PK}_{\text{root}}, Q)$.

A \SPHINCSPLUS{} signature comprises: (i) randomness derived from $\mathsf{SK}_2$, (ii) the \FORS{} signature, and (iii) a sequence of \WOTSPLUS{} signatures with their authentication paths along the hypertree path. Verification reconstructs the \FORS{} public key and validates the \WOTSPLUS{} chain up to $\mathsf{PK}_{\text{root}}$.

For \PHALYNX{}, we preserve this entire structure, but replace the underlying hash function $H$ with Le~Coz et al.'s hash function over $\operatorname{SL}_n(\mathbb{F}_p)$.
This allows us to rely on the security arguments of \SPHINCSPLUS{} which show that the signature scheme remains unforgeable for any hash function that can reasonably be modeled as a random oracle. 

\section{Adapting The Hash Function}
\label{sec:hash}

Le~Coz et al.'s hash function~\cite{LeCoz25pqhash} is based on the Tillich--Z\'emor construction over the special linear group $\operatorname{SL}_n(\mathbb{F}_p)$ \cite{TillichZemor94,TillichZemor08}. The core idea is to interpret an input message as a sequence of symbols that drives a non-backtracking random walk on the Cayley graph of $\operatorname{SL}_n(\mathbb{F}_p)$, with the resulting group element serving as the hash digest.

Existing cryptanalytic results show that group-based hash functions over low-dimensional groups can be vulnerable to efficient attacks.
In particular, Petit, Lauter, and Quisquater~\cite{KL-CP-JQ} demonstrated practical collision and preimage attacks against the LPS and Morgenstern hash families, which are based on $\operatorname{SL}_2(\mathbb{F}_p)$.
These attacks exploit techniques specific to $\operatorname{SL}_2$, and do not directly extend to higher-rank groups.
Following the security recommendations of~\cite{LeCoz25pqhash}, we fix the dimension to $n=4$, as no comparable subexponential attacks are currently known for $\operatorname{SL}_n(\mathbb{F}_p)$ when $n\ge 3$.

\paragraph{Output length and parameterization of the $\mathrm{SL}_4(\mathbb{F}_p)$ hash.}
\label{sec:outputlen}

A \PHALYNX{} digest is obtained by serializing a matrix in $\mathrm{SL}_4(\mathbb{F}_p)$.
An element of $\mathbb{F}_p$ takes values in $\{0,\dots,p-1\}$, hence the minimal number of bits
required to represent it is
\[
\ell_p \;=\; \left\lceil \log_2(p) \right\rceil
\;=\; \left\lfloor \log_2(p-1) \right\rfloor + 1.
\]

\paragraph{Our default 512-bit instantiation.}
We fix $n=4$ and opt for the prime number 
\(
p = 2147483647 = 2^{31}-1
\),
so arithmetic is performed in the prime field $\mathbb{F}_p$ and matrices lie in $\mathrm{SL}_4(\mathbb{F}_p)$.
The minimal representation length is $\ell_p = \lceil \log_2(p)\rceil = 31$ bits.
However, we serialize each matrix entry as a fixed \emph{32-bit unsigned integer} (little-endian),
with the canonical constraint that the decoded value must lie in $\{0,\dots,p-1\}$.
With this fixed-width rule, a $4\times 4$ matrix yields exactly
\(
16 \times 32 = 512 \text{ bits } = 64 \text{ bytes}
\),
which is the digest size used throughout \PHALYNX{}.

This choice intentionally trades a small amount of encoding slack (since $p<2^{32}$) for a simple,
constant-length, word-aligned interface. It also provides a large output space ($2^{512}$ possible digests)
and integrates cleanly with the \SPHINCSPLUS{} design, where a single
fixed output length is consumed by \WOTSPLUS{}, \FORS{}, and the hypertree.

\paragraph{Other output sizes.}
If one targets a digest length $L$, one can either (i) keep $n=4$ and change the fixed-width per-entry encoding,
or (ii) change $(n,p)$ so that the chosen fixed-width packing yields exactly $L$ bits.
In all cases, we require $p$ to be prime (so $\mathbb{F}_p$ is a field).

We now describe the input encoding and the non-backtracking walk that define the hash $\mathcal{G}$.

\paragraph{Input encoding: binary to ternary.} 
\SPHINCSPLUS{} operates on byte strings, but the Tillich--Z\'emor walk requires symbols from a ternary alphabet $\{1,2,3\}$. To bridge this gap efficiently, we convert each input byte into a base-3 representation using a precomputed lookup table. Specifically, each byte $b \in \{0,\dots,255\}$ is uniquely represented in base~3 using 6~\emph{trits} (since $3^6 = 729 > 256$). We precompute a table $\texttt{byte\_to\_trits}[256][6]$ such that
\[
b = \sum_{j=0}^{5} \texttt{trits}[j] \cdot 3^j, \quad \texttt{trits}[j] \in \{0,1,2\}.
\]

However, rather than using all 6 trits unconditionally, we suppress leading zeros in the ternary expansion of each byte. 
This optimization serves two purposes: 
it reduces the expected walk length (improving computational performance); and furthermore, 
it avoids bias toward short walks for small-byte values (e.g., \texttt{0x00} would otherwise produce six 0-trits, mapping to six identical symbols). 

After zero suppression, each remaining trit $t \in \{0,1,2\}$ is mapped to $t+1 \in \{1,2,3\}$ to align with the walk symbol set. The resulting ternary string drives the non-backtracking walk.
\paragraph{Non-backtracking walk.}
We fix two integer matrices $\tilde{A}, \tilde{B} \in \operatorname{SL}_4(\mathbb{Z})$ whose reductions modulo $p$ generate a free subgroup of $\operatorname{SL}_4(\mathbb{F}_p)$. Following \cite{LeCoz25pqhash}, we use the 10th powers $A = \tilde{A}^{10}$ and $B = \tilde{B}^{10}$ to ensure rapid mixing and a large girth in the associated Cayley graph. The walk operates over the four generators $\{A, B, A^{-1}, B^{-1}\}$. To prevent trivial reductions, the walk is constrained to be \emph{non-backtracking}: no step may immediately apply the inverse of the previous step. 

Given a ternary input sequence $(x_1, \dots, x_k) \in \{1,2,3\}^k$, the first symbol $x_1$ is mapped to an initial generator via a fixed bijection $s_0 : \{1,2,3\} \to \{A, B, A^{-1}\}$ (omitting $B^{-1}$ for symmetry). For each subsequent symbol $x_i$ ($i \geq 2$), the generator $g_i$ is selected using a context-dependent map $s_{\lambda} : \{1,2,3\} \to \{A^{\pm1}, B^{\pm1}\} \setminus \{g_{i-1}^{-1}\}$, where $\lambda$ encodes the identity of the previous generator $g_{i-1}$. This ensures the walk remains non-backtracking. The hash value is the matrix product
\[
\mathcal{G}(M) = g_1 g_2 \cdots g_k \mod p \quad \in \operatorname{SL}_4(\mathbb{F}_p).
\]

\begin{figure}[tb]
    \centering
    \includegraphics[width=\textwidth]{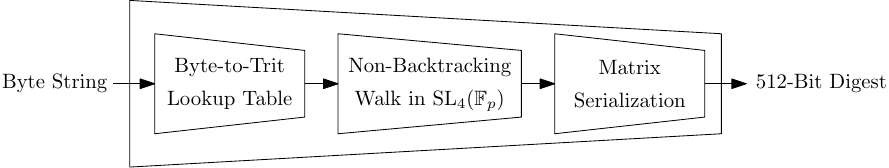}
    \caption{Overview of the \PHALYNX{} Hash Function Pipeline}
    \label{fig:phalynx-hash-pipeline}
\end{figure}

The matrix is then serialized in row-major order, with each entry encoded as a 32-bit little-endian integer, yielding a fixed 512-bit output. 
The \PHALYNX{} hash function pipeline is shown in Fig.~\ref{fig:phalynx-hash-pipeline}. 
The resulting digest is used directly in all \SPHINCSPLUS{} components (\WOTSPLUS{}, \FORS{}, hypertree), with the 256-bit ADRS serving as additional input via domain separation (as detailed in Appendix~\ref{sec:adrs}).

\section{Statistical Evaluation of The \PHALYNX{} Hash Function}
\label{sec:hash-evaluation}
Le~Coz et al.\ provide theoretical arguments for the security of their proposed hash function in $\operatorname{SL}_n(\mathbb{F}_p)$ when $n\ge 3$. Specifically, they show that their proposed hash function provides provable collision and preimage resistance. Furthermore, they provide a theoretical argument that the non-backtracking walk at the heart of the hash function computation is indistinguishable from a random walk for sufficiently long messages, which shows that their hash function behaves like a random oracle for sufficiently long messages. These security arguments apply to the \PHALYNX{} hash function as well, since we only apply one-to-one encodings to the input and output of the original hash function. In this section, we test the \PHALYNX{} hash function empirically to gain practical confidence in its security. 

\paragraph{Experimental Setup.}
To empirically assess the statistical quality of the \PHALYNX{} hash output, we conducted a standard 
NIST Statistical Test Suite (STS~2.1.2)~\cite{nist2010sts} evaluation over $100$ independent binary streams of 
$1{,}000{,}000$ bits each, generated directly from our C reference implementation of the 
$SL_4(\mathbb{F}_p)$--based \PHALYNX{} hash described in Section~\ref{sec:hash}. 

Each stream was produced by hashing successive 16-byte inputs formed from a concatenation of the stream index 
and a 64-bit counter, i.e. 
\(
\texttt{msg} = \texttt{stream\_id} \,\|\, \texttt{ctr}
\).
The generator code (``\texttt{gen\_sts.c}'') was compiled with GCC under Ubuntu~24.04 and produced a 
single file \texttt{sts\_input.bin} of $12.5$~MB (100~streams $\times$ 1{,}000{,}000~bits).
This dataset was supplied as input to the official NIST~STS binary (\texttt{assess~1000000}) using the 
``user-prescribed input file'' mode.

\paragraph{Results and Interpretation.}
The full summary report 
was automatically generated by STS after the successful completion of the statisitcal testing. 
The \PHALYNX{} hash output passed all 15 statistical categories of the NIST~STS suite. 


The key statistics are summarized in Table~\ref{tab:nist-sts}, which reports the global $p$-values and 
the proportion of passing sequences across the 100~streams. 
A sequence is considered to pass if its $p$-value satisfies $p \geq \alpha = 0.01$ and the proportion 
of passes lies within the NIST confidence interval $(0.96,\,1.00)$ for $n=100$. 

\begin{table}[t]
\centering
\caption{NIST SP 800-22 Statistical Test Suite results for \PHALYNX{} hash output 
(100 sequences of 1{,}000{,}000 bits each). 
For tests with two variants (Cumulative Sums, Serial, Random Excursions), 
the first and second $p$-values correspond to the two standard subtests defined in the NIST specification. 
Ranges in the Non-Overlapping Templates test indicate min–max values across all subtests.}

\label{tab:nist-sts}
\begin{tabular}{l@{\quad\quad}cc}
\toprule
\textbf{Statistical Test} & \textbf{$p$-value} & \textbf{Proportion Passed} \\
\midrule
Frequency                      & 0.202268 & 0.99 \\
Block Frequency                & 0.897763 & 0.99 \\
Cumulative Sums (forward\,/\,back) & 0.474986 / 0.016717 & 0.98 / 0.99 \\
Runs                           & 0.595549 & 1.00 \\
Longest Run                    & 0.779188 & 1.00 \\
Rank                           & 0.162606 & 0.98 \\
FFT                            & 0.978072 & 1.00 \\
Non-Overlapping Templates      & 0.003996--0.988549 & 0.96--1.00 \\
Overlapping Template           & 0.494392 & 0.99 \\
Universal                      & 0.699313 & 0.97 \\
Approximate Entropy            & 0.334538 & 0.99 \\
Random Excursions\,/\,Variant    & 0.018969 / 0.988549 & 0.98 / 1.00 \\
Serial (first\,/\,second $p$-values )                 & 0.595549 / 0.719747 & 0.99 / 1.00 \\
Linear Complexity              & 0.249284 & 0.98 \\
\bottomrule
\end{tabular}
\end{table}

The NIST STS results indicate that the \PHALYNX{} hash output exhibits strong statistical randomness across nearly all standard criteria. With the exception of a single instance in the Non-Overlapping Templates test (minimum $p$-value = 0.003996, slightly below the $\alpha = 0.01$ threshold), all other subtests produced $p$-values $\geq 0.01$, and the proportion of passing sequences for every test category lies within the NIST-recommended confidence interval $[0.960, 1.000]$ for 100 sequences.

Notably, complex pattern-detection tests, (including Random Excursions, Random Excursions Variant, Serial, and Approximate Entropy) all achieved excellent pass rates with well-distributed $p$-values, suggesting robust avalanche and diffusion properties in the underlying $SL_4(\mathbb{F}_p)$ walk and modular reduction.

Such minor deviations are not uncommon in large-scale randomness testing: the Non-Overlapping Templates test evaluates 148 distinct bit patterns, and under the null hypothesis of randomness, approximately 1–2\% of individual template tests are expected to fall below $\alpha = 0.01$ purely by chance. Thus, this isolated borderline result is unlikely to indicate a structural weakness.

Overall, the empirical evidence supports the view that the \PHALYNX{} hash function behaves like a statistically uniform pseudorandom generator under the tested conditions, indicating its suitability as a drop-in replacement for standard hash functions in \SPHINCSPLUS{}.

\section{\PHALYNX{}: Design and Integration}
\label{sec:phalynx-design}

\PHALYNX{} is a structurally faithful variant of \SPHINCSPLUS{}, differing in exactly one component: the underlying hash function. We preserve the complete \SPHINCSPLUS{} architecture, including 
\WOTSPLUS{} for leaf authentication, 
\FORS{} as the few-time signature layer,
the layered hypertree construction for stateless many-time signing,
the two-seed key generation (\texttt{SK.seed}, \texttt{SK.prf}),
the 256-bit structured address (\texttt{ADRS}) for domain separation, and
the full signing/verification workflow. 

No modifications are made to the hypertree height, \FORS{} parameters, \WOTSPLUS{} chaining, or signature format. This ensures that \PHALYNX{} inherits the provable security and stateless design of \SPHINCSPLUS{} without alteration.

The sole change is the replacement of the standard hash primitive (e.g., \SHAKE{}) with the \PHALYNX{} hash function specified in Section~\ref{sec:hash}. Concretely, every call to the tweakable hash function
\(
\operatorname{Th}(P, T, M) = H(P \,\|\, T \,\|\, M)
\)
is instantiated by passing the byte string $B = P \,\|\, T \,\|\, M$ to the \PHALYNX{} hash function. The input $B$ is processed byte-by-byte via a precomputed lookup table to produce a ternary sequence, which drives a non-backtracking walk in $\operatorname{SL}_4(\mathbb{F}_p)$. The resulting matrix is serialized to exactly 512 bits.

Our parameter choices, matrix dimension $n = 4$, prime $p = 2^{31} - 1$, and generator exponent $\ell = 10$, are driven by a balance of security against known attacks (e.g., avoiding $\mathrm{SL}_2$-based vulnerabilities~\cite{KL-CP-JQ}) and output compatibility. 
We further explore the parameter space for performance trade-offs in the following. 


\subsection{Exploring The Parameter Space for \PHALYNX{}}
\label{sec:parameters}

Our objective is to achieve an effective collision resistance of approximately 256 bits. By the birthday bound, this requires a 512-bit digest, i.e., $n = 64$ bytes. All parameter sets in this work are therefore instantiated with $n = 64$, using the official \SPHINCSPLUS{} Sage parameter-generation script\footnote{\url{https://sphincs.org/software.html}} configured for this output length.

The Complete list of admissible parameter sets is obtained from this search is reported in Appendix~\ref{ann:param}.  
From this pool, we extract two representative subsets in order to study how different design objectives affect security degradation under signature exposure. 
In particular, we focus on two complementary criteria that are relevant for practical deployments: signing speed and signature size. 
Tables~\ref{tab:candidates_high_speed} and \ref{tab:candidates_low_size} list the top ten parameter sets yielding the highest signing speeds and the smallest signature sizes, respectively.
%



\begin{table}[t]
\centering
\setlength{\tabcolsep}{6pt} 

\begin{minipage}[t]{0.48\linewidth}
\centering
\begin{tabular}{lrrrrr}
\toprule
ID & $h$ & $d$ & $b$ & $k$ & $w$ \\
\midrule
F1 & 64 & 8 & 14 & 22 & 256 \\
F2 & 72 & 9 & 16 & 16 & 256 \\
F3 & 72 & 9 & 17 & 15 & 256 \\
F4 & 63 & 7 & 10 & 38 & 256 \\
F5 & 63 & 7 & 14 & 23 & 256 \\
F6 & 72 & 8 &  7 & 45 & 256 \\
F7 & 72 & 8 & 8 & 37 & 256 \\
F8 & 72 & 8 & 16 & 16 & 256 \\
F9 & 72 & 8 & 17 & 15 & 256 \\
F10 & 72 & 9 & 21 & 12 & 256 \\

\bottomrule
\end{tabular}
\subcaption{Highest signing speed}
\label{tab:candidates_high_speed}
\end{minipage}
\hfill
\begin{minipage}[t]{0.48\linewidth}
\centering
\begin{tabular}{lrrrrr}
\toprule
ID & $h$ & $d$ & $b$ & $k$ & $w$ \\
\midrule
P1 & 72 & 4 & 17 & 15 & 256 \\
P2 & 72 & 4 & 16 & 16 & 256 \\
P3 & 72 & 4 & 21 & 12 & 256 \\
P4 & 80 & 4 &  9 & 28 & 256 \\
P5 & 76 & 4 & 10 & 26 & 256 \\
P6 & 80 & 4 &  7 & 38 & 256 \\
P7 & 64 & 4 & 14 & 22 & 256 \\
P8 & 75 & 5 & 19 & 13 & 256 \\
P9 & 80 & 5 &  9 & 28 & 256 \\
P10& 75 & 5 & 14 & 18 & 256 \\
\bottomrule
\end{tabular}
\subcaption{Lowest signature size}
\label{tab:candidates_low_size}
\end{minipage}
\hfill

\caption{Candidate parameter sets for \PHALYNX{}.}
\label{tab:all_candidates}
\end{table}

The parameters in Tables~\ref{tab:candidates_high_speed} and \ref{tab:candidates_low_size} are used in the next section to analyze FORS-based security degradation as a function of the number of issued signatures.
This analysis allows us to compare parameter sets across different optimization goals and to identify recommended \PHALYNX{} instances that balance signature speed and size under signature exposure.

\subsection{\FORS{}-Based Security Degradation Under Signature Exposure}
\label{sec:fors-degradation}


In stateful and stateless hash-based signature schemes, repeated use of a single secret key inevitably reveals partial secret information through published signatures. 
In \SPHINCSPLUS{}, this phenomenon is most prominent in \FORS{} component, where each signature discloses exactly one secret value from each of $k$ disjoint subsets.
As the number of issued signatures grows, the probability that an adversary can recombine previously revealed values into a valid forgery increases.

The purpose of this section is to quantify and visualize the resulting \emph{security degradation} as a function of the number of signatures issued under a single \SPHINCSPLUS{} key.
Our analysis follows the explicit probabilistic model used in \cite{KreyviumSphincsSmaller}.


We consider the \FORS{} forgery attack described by Fluhrer and Dang~\cite{KreyviumSphincsSmaller}.
Let $2^m$ be the number of signatures available to the adversary.
Each signature independently selects a leaf in the hypertree, and the probability that a given signature maps to the same \FORS{} instance as a fixed target message is $2^{-h}$, where $h$ is the hypertree height.
Consequently, the number $G$ of signatures colliding on the target \FORS{} follows a binomial distribution
\[
G \sim \mathrm{Binomial}(2^m, 2^{-h}),
\]
which is well approximated by a Poisson distribution with mean $\lambda = 2^{m-h}$. 

Conditioned on having exactly $g$ signatures on the same \FORS{} instance, each \FORS{} tree reveals one element out of $t = 2^b$ possible leaves.
For a fixed tree, the probability that none of the $g$ signatures reveals the target leaf is $(1 - t^{-1})^g$, and thus the probability of at least one hit is $1 - (1 - t^{-1})^g$. 
Since a valid forgery requires this condition to hold simultaneously for all $k$ \FORS{} trees, the conditional success probability is $\left(1 - (1 - t^{-1})^g\right)^k$. 
Applying the law of total probability over the Poisson-distributed random variable $G$, we obtain the forgery probability
\begin{equation}
\label{eq:fors-poisson}
p(m) = \sum_{g=0}^{\infty} \frac{\lambda^g e^{-\lambda}}{g!}
\left(1 - (1 - t^{-1})^g\right)^k,
\quad \text{where} \quad 
\lambda = 2^{m-h}.
\end{equation}
Equation~\eqref{eq:fors-poisson} corresponds to the \FORS{} recombination model used in \cite{KreyviumSphincsSmaller}.


To interpret $p(m)$ in cryptographic terms, we express it as an \emph{effective security level}, i.e. $\mathrm{Sec}(m) = -\log_2 p(m)$, which represents the number of bits of security associated with the \FORS{} component after $2^m$ signatures.

To isolate the \emph{degradation} caused by signature exposure, we normalize with respect to a baseline number of signatures $2^{m_0}$:
\begin{equation}
\label{eq:delta-sec}
\Delta \mathrm{Sec}(m) = \max\{0, \mathrm{Sec}(m_0) - \mathrm{Sec}(m)\}, 
\quad \text{where} \quad 
\mathrm{Sec}(m) = -\log_2 p(m)
\end{equation}
This measures the loss attributable solely to accumulation of signatures, independently of other attack avenues (e.g., hash preimage attacks).

Finally, we define the exposed-key effective security as
\begin{equation}
\label{eq:sec-eff}
\mathrm{Sec}_{\mathrm{eff}}(m) = \mathrm{Sec}_{\mathrm{target}} - \Delta \mathrm{Sec}(m),
\end{equation}
where $\mathrm{Sec}_{\mathrm{target}}$ is the nominal level (e.g., 256 bits).

\begin{figure}
  \centering
  \includegraphics[width=0.98\textwidth]{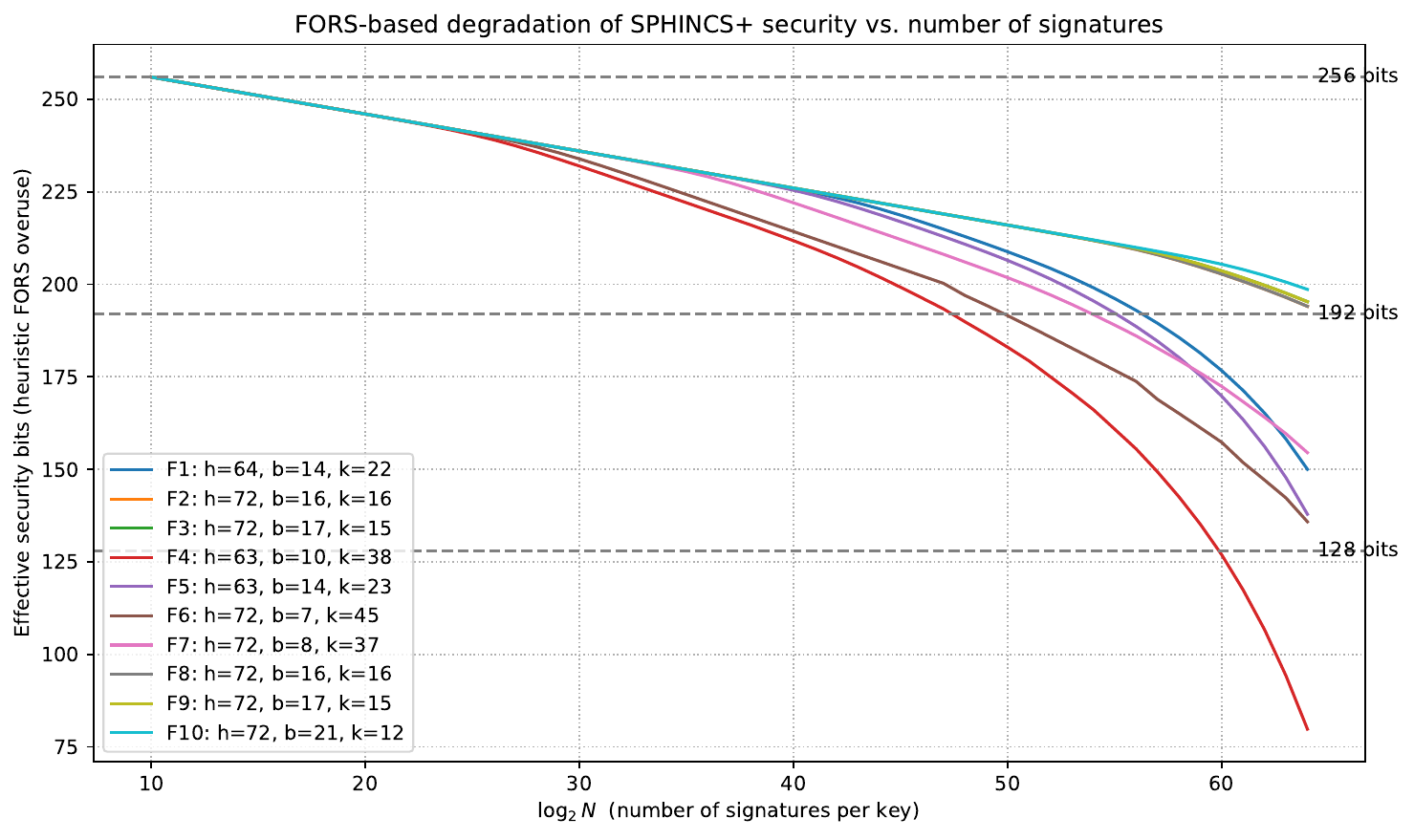}
  \caption{\FORS{}-based effective security degradation versus the signature budget $2^m$ for parameter sets selected by low signing cost. 
  }
  \label{fig:fast}
\end{figure}
\begin{figure}
  \centering
  \includegraphics[width=0.98\textwidth]{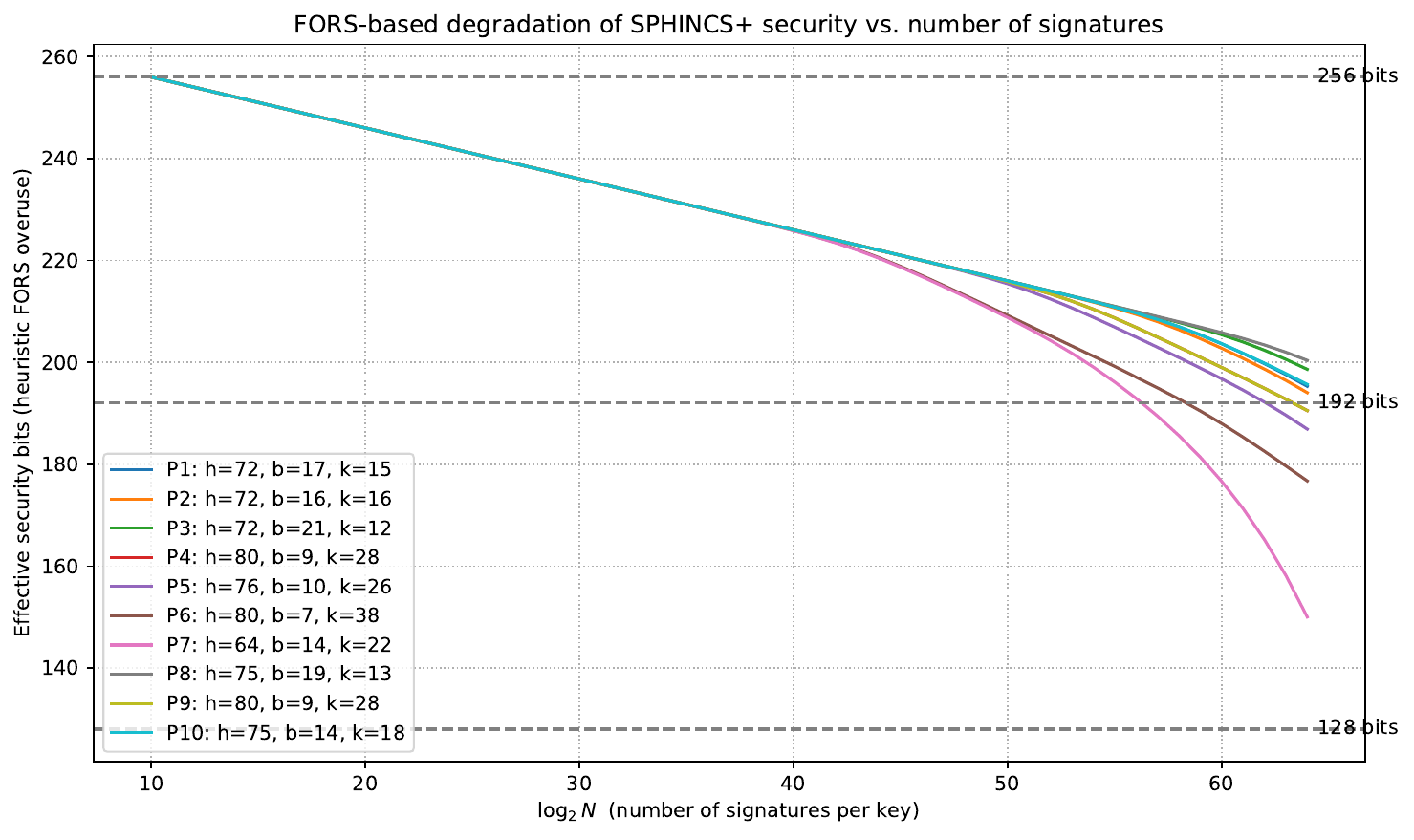}
  \caption{\FORS{}-based effective security degradation versus the signature budget $2^m$ for parameter sets selected by minimal signature size. 
  }
  \label{fig:small}
\end{figure}

Using the model in Equations~\eqref{eq:fors-poisson}, \eqref{eq:delta-sec}, and \eqref{eq:sec-eff}, we evaluate $\mathrm{Sec}_{\mathrm{eff}}(m)$ for the two families of parameter sets for highest signing speeds and lowest signature sizes specified in the previous section. 
For each set $(h,b,k)$, we compute the degradation curve over a wide range of $m$ (i.e., over a wide range of signature budgets $2^m$).
Figures~\ref{fig:fast} and \ref{fig:small} report the degradation for the ten parameter sets in each of the families. 


Across both families, the security degradation accelerates as the number of signatures increases.
In the early regime, the expected number of collisions on a target FORS instance remains small because $\lambda = 2^{m-h}$ is small; consequently, the Poisson mass concentrates on small $g$ and the forgery probability $p(m)$ stays negligible.
As $m$ grows, $\lambda$ increases and the probability of observing larger collision multiplicities $g$ rises sharply, which amplifies the coverage term $\left(1-(1-t^{-1})^g\right)^k$ and accelerates the degradation.

Comparing Figures~\ref{fig:fast} and \ref{fig:small}, we observe that parameter sets selected for minimal signature size tend to degrade faster than parameter sets selected for low signing cost.
This behavior is consistent with the structure of Equation~\eqref{eq:fors-poisson}: increasing $k$ strengthens the requirement of simultaneously covering all $k$ subsets and makes the attack probability more sensitive to repeated exposures, while smaller $b$ (hence smaller $t=2^b$) increases the hit probability per trial.
In contrast, parameter sets associated with lower signing cost typically employ \FORS{} parameters that delay the onset of the rapid-growth regime, yielding slower degradation over comparable signature budgets.

\subsection{Selecting A \PHALYNX{} Parameter Set}
\label{sec:paramselect}
Depending on the use case, one can select a parameter set for \spinel\ that balances three constraints: 
(i) strong effective security under \FORS{} exposure when many signatures are issued under one long-term key, 
(ii) acceptable signature size, and 
(iii) measurable performance under a fixed benchmarking methodology. The key point is that \FORS{} exposure is \emph{usage-dependent} (as discussed in the previous section): as more signatures are published under the same public key, more \FORS{} secret values are revealed, and the probability of recombining previously revealed material into a forgery increases. Therefore, parameter selection must be tied to an explicit operational assumption on how many signatures a key is expected to produce.

We model this assumption by fixing an operational signature budget \(2^{m^\star}\), where \(m^\star\) is the logarithm in base 2 of the intended maximum number of signatures issued under a single long-term public key during its lifetime. For each candidate parameter set \((h,d,b,k)\), one can compute the \FORS{} exposure curve \(\mathrm{Sec}_{\mathrm{eff}}(m)\) using Equations~\eqref{eq:fors-poisson}, \eqref{eq:delta-sec}, and \eqref{eq:sec-eff}, and record the effective security at the budget point, \(\mathrm{Sec}_{\mathrm{eff}}(m^\star)\). Then, only candidates meeting a minimum exposure threshold \(\mathrm{Sec}_{\mathrm{eff}}(m^\star)\ge \sigma\) are retained, where \(\sigma\) is a conservative target (e.g., \(\sigma\in\{192,200\}\) bits). Among the remaining candidates, the instance achieving the best size/performance trade-off (i.e., signature cost and length) is finally selected subject to the exposure constraint.


A final consideration for parameter selection is the observation that the exposure curves in Figures~\ref{fig:fast} and \ref{fig:small} show a consistent separation between candidates: instances with larger \(b\) (hence larger \FORS{} leaf space) tend to delay degradation under large signature budgets, while moderate \(k\) avoids unnecessary overhead. 


\section{Performance Evaluation}
\label{sec:performance}

To evaluate the practical efficiency of \PHALYNX{}, we implemented the scheme in~C and benchmarked its performance for the parameter sets achieving the best signing speeds, as listed in Table~\ref{tab:candidates_high_speed}, and measuring the same metrics used in benchamrking \SPHINCSPLUS{}~\cite{BernsteinHKNRS19sphincs+}. 

Our implementation builds upon the reference codebase of the \SPHINCSPLUS{} signature scheme, as provided in the official NIST PQC submission repository\footnote{\url{https://github.com/sphincs/sphincsplus}}. We adapted this code to integrate our proposed hash function and to enable performance comparison against the original instantiation. All modifications were designed to preserve the structural and security properties of the base scheme while substituting the underlying hash primitive. The complete source code, including build scripts and benchmarking utilities, will be released publicly upon publication to ensure reproducibility and community verification.

The implementation uses the \spinel\ hash function (described in Section~\ref{sec:hash}) with output length 512\, ($n = 64$) and Winternitz parameter $w = 256$.
We evaluated ten parameter sets that all target 128-bit post-quantum security, while varying the hypertree height $h$, the \FORS{} tree height $b$, and the number of \FORS{} trees $k$. 
To enable comparison between different specifications, the costs of key generation, signing, and verification were measured in the number of clock cycles. 
Cycle counts were measured using the processor timestamp counter, and each data point reported below is the \emph{median} over $10$ independent runs to reduce the impact of system noise. 

To ensure stable and reproducible cycle counts, all benchmarks were executed on 
the \emph{Viking} high-performance computing cluster at the University of York. 
Each node is equipped with two AMD EPYC 7643 processors (48 cores each), and measurements were performed on a single core with CPU frequency scaling disabled. 
All benchmarks were executed on the same compute node with identical compilation and runtime settings, ensuring fair comparison across parameter sets.
The results are given in Table~\ref{tab:phalynx-performance-viking}.

\begin{table}[t]
  \centering
  \caption{Performance of \PHALYNX{} on an AMD EPYC 7643 48-Core Processor @ 2295.717\,MHz
  (medians over 10 runs). In all cases: $n=64$ and $w=256$.}
  \label{tab:phalynx-performance-viking}
  \begin{tabular}{l@{~~~}r@{~~~}r@{~~~}r@{~~~}r@{~~~}r@{~~~}r}
    \toprule
    Parameters &
      \multicolumn{3}{c}{Cycles (median)} &
      \multicolumn{3}{c}{Bytes} \\
    \cmidrule(lr){2-4}\cmidrule(lr){5-7}
    $(h,d,b,k)$ & keypair & sign & verify & sig & pk & sk \\
    \midrule
    $(64, 8, 14, 22)$
      & 266\,889\,736\,639
      & 2\,183\,968\,488\,295
      & 6\,355\,211\,466
      & 59\,072
      & 128
      & 256 \\
    $(72, 9, 16, 16)$
      & 275\,117\,652\,288
      & 2\,646\,740\,656\,479
      & 7\,114\,693\,442
      & 60\,096
      & 128
      & 256 \\
    $(72, 9, 17, 15)$
      & 277\,270\,999\,860
      & 2\,831\,860\,077\,991
      & 5\,709\,825\,003
      & 59\,968
      & 128
      & 256 \\
    $(63, 7, 10, 38)$
      & 558\,916\,025\,894
      & 3\,743\,229\,415\,977
      & 4\,941\,522\,066
      & 60\,416
      & 128
      & 256 \\
    $(63, 7, 14, 23)$
      & 573\,357\,992\,211
      & 3\,931\,649\,658\,362
      & 4\,946\,136\,062
      & 55\,744
      & 128
      & 256 \\
    $(72, 8, 7, 45)$
      & 542\,221\,601\,750
      & 4\,438\,379\,403\,753
      & 6\,894\,332\,834
      & 61\,504
      & 128
      & 256 \\
    $(72, 8, 8, 37)$
      & 523\,680\,116\,652
      & 4\,044\,495\,613\,372
      & 5\,509\,524\,443
      & 59\,776
      & 128
      & 256 \\
    $(72, 8, 16, 16)$
      & 522\,234\,093\,047
      & 4\,170\,075\,435\,064
      & 5\,476\,275\,953
      & 55\,872
      & 128
      & 256 \\
    
    $(72, 8, 17, 15)$
      & 526\,713\,225\,483
      & 4\,422\,912\,786\,585
      & 5\,293\,148\,861
      & 55\,744
      & 128
      & 256 \\
    
    $(72, 9, 21, 12)$
      & 258\,695\,804\,893
      & 5\,348\,844\,423\,749
      & 6\,753\,064\,776
      & 59\,584
      & 128
      & 256 \\
    \bottomrule
  \end{tabular}
\end{table}

Because \PHALYNX{} inherits the deterministic hash-call structure of \SPHINCSPLUS{}, the total cost of key generation, signing, and verification is proportional to the cost of a single hash  (\texttt{thash}) and to the fixed number of hash calls required by each operation. In our implementation, a single \texttt{thash} call takes in the order of $10^5$ cycles, which matches the measured end-to-end costs reported here.

To gain further confidence in our benchmarks, we also tested \SPHINCSPLUS{} under an identical setting. The cycle count results under various hash function configurations largely agree with those reported by the designer of \SPHINCSPLUS{}~\cite{BernsteinHKNRS19sphincs+}, confirming the soundness of our measurement method. As these results may be of independent interest, they are given in Appendix~\ref{ann:sphincs-benchmarks}. 


As Table~\ref{tab:phalynx-performance-viking} shows, signing is consistently more computationally expensive than verification across all parameter sets. While verification costs in the order of $10^{9}$ cycles, signing consistently requires in the order of $10^{12}$ cycles, which makes signing three orders of magnitude slower than verification. This shows that \spinel\ is better suited for applications in which signing takes place less regularly than verification or is performed by entities that have access to more computational resources than those who carry out verification. An example of such applications is public-key certificates. 

Regarding parameter selection, the results show that increasing the hypertree height $h$ increases the signing cost as expected. 
For instance, comparing parameter sets $(64,8,14,22)$ and $(72,9,16,16)$, the signing cost increases by around $20\%$, consistent with longer authentication paths and more hash work overall. 
Furthermore, one can see that within the same $(h,d)$ regime, \FORS{} parameters mainly drive the signing--verification trade-off. 
For example, for $(h,d)=(72,9)$, switching from $(b,k)=(16,16)$ to $(17,15)$ increases the signing cost (from $2.65\times 10^{12}$ to $2.83\times 10^{12}$ cycles) while reducing the verification cost (from $7.11\times 10^{9}$ to $5.71\times 10^{9}$ cycles). This illustrates that modest changes in \FORS{} structure can noticeably impact the verification cost even when key sizes are unchanged. 

Finally, note that, even when limiting our attention to a select number of parameter sets, the bandwidth--compute trade-off is clearly visible. 
For example, for $(h,d)=(63,7)$, moving from $(b,k)=(10,38)$ to $(14,23)$ reduces the signature size from around 60\,KB to 56\,KB (an around $7\%$ reduction) at the expense of higher key generation and signing costs (both increasing by around $3$--$5\%$), while verification cost remains essentially unchanged.

As expected, \PHALYNX{} exhibits larger signature sizes and slower verification times compared to \SPHINCSPLUS{} instantiated with \SHAKE{}, due to the computational overhead of matrix operations in $\operatorname{SL}_4(\mathbb{F}_p)$ and the use of a 512-bit digest. This trade-off is inherent: stronger algebraic hardness assumptions and higher security levels naturally incur performance and bandwidth costs. Nevertheless, the scheme remains functionally viable, and future optimizations, such as signature compression techniques analogous to those in \SPHINCSPLUS{}\texttt{C}~\cite{SphincsPlusCompressed} or structured Merkle tree variants, could significantly reduce its computational costs without compromising security.

\section*{Conclusion}
\label{sec:conclusion}

We introduced \PHALYNX{}, a stateless hash-based signature scheme that preserves the full \SPHINCSPLUS{} structure while replacing its underlying hash primitive with a Tillich--Z\'emor Cayley-walk hash over $\mathrm{SL}_4(\mathbb{F}_p)$. The design keeps the standard \WOTSPLUS{}, \FORS{}, and hypertree composition unchanged and instantiates all internal oracles through a single domain-separated base primitive, enabling the preservation of formal security guarantees in the post-quantum setting. 

We specified a concrete 512-bit instantiation with fixed-width serialization, and empirically verified the statistical randomness behavior of the hash function's output using the NIST STS test suite. After demonstrating how this adapted hash function can be integrated into the \SPHINCSPLUS\ framework, we quantified its \FORS{}-driven security degradation under signature exposure using the exact Poisson model from prior work and provided degradation curves for representative parameter families, supporting informed parameter selection under realistic signature budgets. Finally, we implemented \PHALYNX{} and reported reproducible benchmarks for computation cost and signature size, confirming that the construction is practical in security-critical scenarios where signatures are generated infrequently, such as for public-key certificates. 

We envisage at least two directions for future work. \spinel\ could be optimized using approaches from the literature such as signature compression techniques. Besides, further exploration of the hash function design space could provide a more diverse set of security assumptions and parameters, thereby enabling a wider range of choices for different applications. 

\begin{credits}
\subsubsection{\ackname} 
The authors are grateful to Ludovic Perret for valuable discussions, proposing the problem and insightful feedback during the early stages of this work. 
The authors acknowledge the support from the Institut Henri Poincaré (UAR 839 CNRS-Sorbonne Université) and LabEx CARMIN (ANR-10-LABX-59-01). 
This project started during the Trimester on Post-quantum Algebraic Cryptography at IHP. 
DK conducted this work partially with the support of ONR Grant 62909-24-1-2002. 
DK also thanks Institut des Hautes \'Etudes Scientifiques (IHES) for providing stimulating environment while this project was partially done. 
DK was supported by the CARMIN fellowship during the completion of this project. 
The Viking cluster was used during this project, which is a high performance compute facility provided by the University of York. We are grateful for computational support from the University of York IT Services and the Research IT team. 

\end{credits}

%
%

\appendix

\section{\SPHINCSPLUS{} Address Structure}
\label{sec:adrs}

To ensure domain separation across all hash invocations, \SPHINCSPLUS{} employs a structured 256-bit address (\texttt{ADRS}), partitioned into eight 32-bit words. \texttt{ADRS} encodes the exact context of each hash call, guaranteeing that identical inputs in different roles (e.g., \WOTSPLUS{} chaining vs. Merkle node hashing) produce independent outputs.

\noindent
The address comprises four main fields:
\begin{itemize}

    \item \texttt{layer}: the hypertree layer (1 word),
    \item \texttt{tree}: the tree index within the layer (3 words),
    \item \texttt{type}: the operation type (1 word), and 
    \item \texttt{context}: type-dependent parameters (3 words).
\end{itemize}

\noindent
Five \texttt{type} values are defined~\cite{SphincsPlus}:
\begin{enumerate}
    \item \WOTSPLUS{} chaining, including key pair index, chain address, and hash iteration,
    \item \WOTSPLUS{} public key compression,
    \item Merkle tree nodes, including node height and index,
    \item \FORS{} tree nodes, including key pair index, internal height, and leaf index, 
    \item \FORS{} root compression.
\end{enumerate}

\texttt{ADRS} serves as the \emph{tweak} in \SPHINCSPLUS{}’s tweakable hash function $\operatorname{Th}$~\cite{BernsteinHKNRS19sphincs+}, which is defined as:
\[
\operatorname{Th}(P, T, M) = H(P \,\|\, T \,\|\, M),
\]
where $P$ is public parameter (e.g., public seed), $T \in \{0,1\}^{256}$ is the \texttt{ADRS} tweak, and $M$ is the message. This construction ensures that all components, \WOTSPLUS{}, \FORS{}, and Merkle layers, operate in disjoint hash domains, a prerequisite for the scheme’s security proofs in the (quantum) random oracle model.

In \PHALYNX{}, we preserve this \texttt{ADRS} structure for our Tillich–Zémor hash over $\operatorname{SL}_n(\mathbb{F}_p)$, ensuring equivalent domain separation.

\section{Exploration of The Parameter Space}
\label{ann:param}
Table~\ref{tbl:parameter-space} presents the \SPHINCSPLUS{}/\PHALYNX{} parameters used to achieve 256-bit security, including signature size (in KB) and signing cost (in billions of hashes, inversely indicating signing speed) for each configuration.

\begin{longtable}{|@{\ }c@{\ }|@{\ }c@{\ }|@{\ }c@{\ }|@{\ }c@{\ }|@{\ }c@{\ }|@{\ }c@{\ }|@{\ }c@{\ }|@{\ }c@{\ }|}
\caption{\spinel's parameter sets calculated for 256-bit security}
\label{tbl:parameter-space} \\ 
\hline
\textbf{h} & \textbf{d} & \textbf{b} & \textbf{k} & \textbf{w} & \textbf{sec} & \begin{tabular}{c} \textbf{Signature Size} \\ \textbf{(KB)} \end{tabular} & \begin{tabular}{c} \textbf{Signing Cost} \\ \textbf{(billions of hashes)} \end{tabular} \\
\hline
\endfirsthead
\hline
\textbf{h} & \textbf{d} & \textbf{b} & \textbf{k} & \textbf{w} & \textbf{sec} & \begin{tabular}{c} \textbf{Signature Size} \\ \textbf{(KB)} \end{tabular} & \begin{tabular}{c} \textbf{Signing Cost} \\ \textbf{(billions of hashes)} \end{tabular} \\
\hline
\endhead
60 & 4 & 13 & 31 & 256 & 256 & 47.5 & 2215.23 \\
60 & 5 & 13 & 31 & 256 & 256 & 52.8 & 346.56 \\
60 & 6 & 13 & 31 & 256 & 256 & 57.0 & 104.32 \\
60 & 4 & 18 & 19 & 16 & 256 & 60.5 & 284.82 \\
60 & 4 & 18 & 19 & 256 & 256 & 43.9 & 2224.69 \\
60 & 5 & 18 & 19 & 256 & 256 & 48.1 & 356.01 \\
60 & 6 & 18 & 19 & 256 & 256 & 52.3 & 113.78 \\
63 & 7 & 10 & 38 & 256 & 256 & 60.4 & 60.64 \\
63 & 7 & 14 & 23 & 256 & 256 & 55.7 & 61.31 \\
64 & 4 & 14 & 22 & 16 & 256 & 58.8 & 550.44 \\
64 & 4 & 14 & 22 & 256 & 256 & 42.0 & 4430.17 \\
64 & 8 & 14 & 22 & 256 & 256 & 59.1 & 35.33 \\
65 & 5 & 8 & 49 & 256 & 256 & 53.6 & 692.13 \\
65 & 5 & 9 & 40 & 256 & 256 & 50.9 & 692.14 \\
66 & 6 & 9 & 38 & 256 & 256 & 54.0 & 207.67 \\
66 & 6 & 15 & 19 & 256 & 256 & 49.1 & 208.88 \\
68 & 4 & 8 & 42 & 16 & 256 & 62.1 & 4397.73 \\
68 & 4 & 8 & 42 & 256 & 256 & 44.6 & 3543.56 \\
70 & 5 & 9 & 33 & 256 & 256 & 46.8 & 1384.24 \\
70 & 7 & 9 & 33 & 256 & 256 & 55.2 & 121.15 \\
72 & 4 & 7 & 45 & 16 & 256 & 61.2 & 2198.88 \\
72 & 4 & 7 & 45 & 256 & 256 & 44.6 & 1771.78 \\
72 & 6 & 7 & 45 & 256 & 256 & 53.0 & 415.27 \\
72 & 8 & 7 & 45 & 256 & 256 & 61.5 & 69.22 \\
72 & 4 & 8 & 37 & 16 & 256 & 59.5 & 2198.88 \\
72 & 4 & 8 & 37 & 256 & 256 & 43.9 & 1771.78 \\
72 & 6 & 8 & 37 & 256 & 256 & 51.3 & 415.28 \\
72 & 8 & 8 & 37 & 256 & 256 & 59.8 & 69.23 \\
72 & 4 & 16 & 16 & 16 & 256 & 55.6 & 2200.96 \\
72 & 4 & 16 & 16 & 256 & 256 & 38.9 & 1771.98 \\
72 & 6 & 16 & 16 & 256 & 256 & 47.4 & 417.36 \\
72 & 8 & 16 & 16 & 256 & 256 & 55.9 & 71.31 \\
72 & 9 & 16 & 16 & 256 & 256 & 60.1 & 41.03 \\
72 & 4 & 17 & 15 & 16 & 256 & 55.4 & 2202.80 \\
72 & 4 & 17 & 15 & 256 & 256 & 38.8 & 1772.17 \\
72 & 6 & 17 & 15 & 256 & 256 & 47.3 & 419.19 \\
72 & 8 & 17 & 15 & 256 & 256 & 55.7 & 73.14 \\
72 & 9 & 17 & 15 & 256 & 256 & 59.9 & 42.86 \\
72 & 4 & 21 & 12 & 16 & 256 & 55.1 & 2249.20 \\
72 & 4 & 21 & 12 & 256 & 256 & 38.5 & 1776.81 \\
72 & 6 & 21 & 12 & 256 & 256 & 46.9 & 465.59 \\
72 & 8 & 21 & 12 & 256 & 256 & 55.4 & 119.54 \\
72 & 9 & 21 & 12 & 256 & 256 & 59.6 & 89.26 \\
75 & 5 & 14 & 18 & 256 & 256 & 43.3 & 2768.99 \\
75 & 5 & 19 & 13 & 16 & 256 & 63.4 & 357.20 \\
75 & 5 & 19 & 13 & 256 & 256 & 42.6 & 2782.04 \\
76 & 4 & 6 & 51 & 16 & 256 & 61.3 & 4397.77 \\
76 & 4 & 6 & 51 & 256 & 256 & 44.7 & 3543.56 \\
76 & 4 & 10 & 26 & 16 & 256 & 56.8 & 4398.77 \\
76 & 4 & 10 & 26 & 256 & 256 & 40.1 & 3543.56 \\
80 & 4 & 7 & 38 & 16 & 256 & 58.2 & 8795.47 \\
80 & 4 & 7 & 38 & 256 & 256 & 41.5 & 7087.12 \\
80 & 5 & 7 & 38 & 256 & 256 & 45.8 & 5536.82 \\
80 & 4 & 9 & 28 & 16 & 256 & 56.6 & 8795.48 \\
80 & 4 & 9 & 28 & 256 & 256 & 40.0 & 7087.12 \\
80 & 5 & 9 & 28 & 256 & 256 & 44.2 & 5536.84 \\
\hline
\end{longtable}

As long as the parameter $n$ is fixed to $512$ bits, the size of the public key is always $2n$ bits (i.e., $128$ bytes), and the size of the secret key is always $4n$ bits (i.e., $256$ bytes), regardless of the other parameter choices in the table.

\section{\SPHINCSPLUS{} Benchamrking Results}
\label{ann:sphincs-benchmarks}
Table~\ref{tab:sphincs-performance-viking} shows the benchmarking results for \SPHINCSPLUS{} using he same hardware and methods as our reported \spinel\ results in Section~\ref{sec:performance}. 



\begin{table}[H]
  \centering
  \caption{Performance of \SPHINCSPLUS{} on an AMD EPYC 7643 48-Core Processor @ 2295.717\,MHz
  (medians over 10 runs).}
  \label{tab:sphincs-performance-viking}
  \begin{tabular}{l@{\qquad}r@{\qquad}r@{\qquad}r}
    \toprule
    Hash Function &
      \multicolumn{3}{c}{Cycles (median)} \\
    \cmidrule(lr){2-4}
    Configuration & keypair & sign & verify \\
    \midrule
        \texttt{haraka-robust} & 564,832,171 & 8,205,224,470 & 12,102,715 \\ 
        \texttt{shake-robust} & 580,875,338 & 6,621,223,842 & 9,961,518 \\ 
        \texttt{sha2-robust} & 505,295,314 & 5,629,341,287 & 8,593,674 \\ 
        \texttt{haraka-simple} & 307,586,820 & 4,697,310,723 & 6,433,548 \\ 
        \texttt{shake-simple} & 292,412,340 & 3,510,597,318 & 4,639,790 \\ 
        \texttt{sha2-simple} & 175,766,322 & 2,200,746,743 & 3,014,173 \\
    \bottomrule
  \end{tabular}
\end{table}

\end{document}